\documentstyle[prl,aps,multicol,epsf]{revtex}

\begin{document}
\draft
\tighten

\title{ Metallic stripes: separation of spin-, charge- and string fluctuation.}
\author{J. Zaanen, O. Y. Osman,  and W. van Saarloos}
\address{Institute Lorentz, Leiden University, P.O. Box 9506, NL-2300 RA Leiden, The Netherlands}
\date{\today}
\maketitle

\begin{abstract}
Inspired by the cuprate stripes, we consider the problem of a one dimensional
metal living on a delocalized trajectory in two dimensional space: the
metallic lattice string. A model is constructed with maximal coupling between
longitudinal and transversal charge motions, which nevertheless renormalizes
into a minimal generalization of the Luttinger liquid: an independent set
of string modes has to be added to the long wavelength theory, with a dynamics 
governed by the quantum sine-Gordon model. 
\end{abstract}


\begin{multicols}{2}
\narrowtext
Evidence is accumulating that the superconducting state in the cuprates
is closely related the stripe phase, where the active charge
degrees of freedom confine on anti-phase
boundaries in the antiferromagnetic background\cite{tranlast}.
 The case can be made that 
the stripes might be internally like one dimensional (1D) 
metals\cite{trancoex,capec}.
Several theoretical works have appeared taking this `self-organized' 
one dimensionality as a starting point\cite{spingappr}. 
However,
compared to conventional one dimensional metals, stripes are at the least
qualitatively different in one regard: {\em the trajectory on which the
metal lives is itself delocalized in space,} obviously so because of the
absence of static striped order in either the superconducting- or normal states.
The question arises what can be said about the general nature of
a quantum string which is internally a metal. According to the Luttinger
liquid theory of 1d metallicity, all what matters at long wavelength are
the collective charge- and spin oscillations which are governed by 
quantum sine-Gordon field theories (QSG)\cite{repphys}. 
It can be argued that the strings of relevance to cuprates
are governed by QSG as well\cite{eskes}. Here we will demonstrate that at
least in principle a fixed point theory exists which is a minimal 
generalization of the Luttinger liquid:  a
metallic string can be like a Luttinger liquid, except that a
set of string modes has to be added for the theory to be complete.
Our construction rests on the assumption that a reference string state
exists which is at the same time localized in space and internally 
charge-incompressible due to a charge density wave (CDW) instability. 
The CDW solitons emerging under doping
 fluctuate at the same time the string position and the resulting
charged kink gas maps on a spin-full fermion problem with a Luttinger
liquid long wavelength regime. 
  
In the context of cuprates, it might be that the state at $x=1/8$ corresponds
with internally insulating, localized stripes. Since the average stripe separation ($d$) decreases like $1/x$ for $x \le 1/8$, the stripes are likely
internally charge incompressible\cite{capec}. 
For $x > 1/8$, $d$ becomes approximately $x$ independent,
suggesting that the stripes are `doped' with the additional
holes. At the same time, the static stripe phase shows a special 
stability at $x=1/8$, and this might reflect a tendency towards
localization on the single stripe level. For modelling purposes we assume
the electronic system on the stripe to be dominated by short range 
repulsive interactions, favoring a $4k_F$ CDW instability as suggested by
 Nayak and Wilczek\cite{nayak} (Fig. 1a)  -- the other 
possibilities\cite{capec} are more complicated, but not necessarily
qualitatively different in the present context. Finally, it is assumed that
spin separates at the very beginning and can be ignored all along. 
Since the stripe sweeps through a spin-full background, the neglect of 
spin is certainly not justifyable, and further work is needed on this
fascinating problem. 

If the string does not delocalize, the remaining problem of a doped $4k_F$
charge density wave is well understood\cite{hal,num4kf}. 
A representative model, in the sense of
adiabatic continuity, is the extended Hubbard model with both $U$ 
(on-site repulsion) and $V$ (nearest-neighbor repulsion) large compared to 
the bandwidth. At low doping lattice commensuration dominates and the 
relevant lattice scale physics is that of {\em solitons}. Using simple
kinematics,  Kivelson and Schrieffer\cite{kivschr} pointed out that the
injected hole splits into propagating soliton- and anti-soliton excitations,
both carrying half the charge of the hole (fig. 1b). The soliton dynamics is
described in terms of a {\em spinless} fermion problem,     
\begin{equation}
\label{cdwsol}
H_{CDW} \; = \; \sum_{ij} t'(ij) c^{\dagger}_i c_j + 
\sum_{ij} V'(ij) n_i n_j
\end{equation}
where $c^{\dagger}$ creates a soliton, subject to short range hoppings ($t'$)
and (repulsive) interactions ($V'$). This problem is dual to a bosonic
quantum sine Gordon theory,
\begin{equation} 
\label{cdwlut}
H_{\rho, ren} = { {v_{\rho}} \over 2} 
\int dx \left[ K_{\rho} \Pi^2_{\rho} + {1 \over K_{\rho}} 
( \partial_x \phi_{\rho} )^2  + g sin (\alpha \phi_{\rho}) \right] \; , 
\end{equation}
 where $v_c$ and $K_c$ correspond with the charge
velocity and charge stiffness, respectively. Away from quarter filling,
this theory is in the weak coupling regime (the sine interaction is
an irrelevant operator) and the long wavelength dynamics is governed
by free field theory (Luttinger liquid),
 completely specified by the renormalized stiffness
and velocity. These parameters have to be calculated numerically, and
their behavior is well documented for the extended Hubbard 
model\cite{hal,num4kf}.

\begin{figure}[h]
\hspace{0.15 \hsize}
\epsfxsize=0.6\hsize
\epsffile{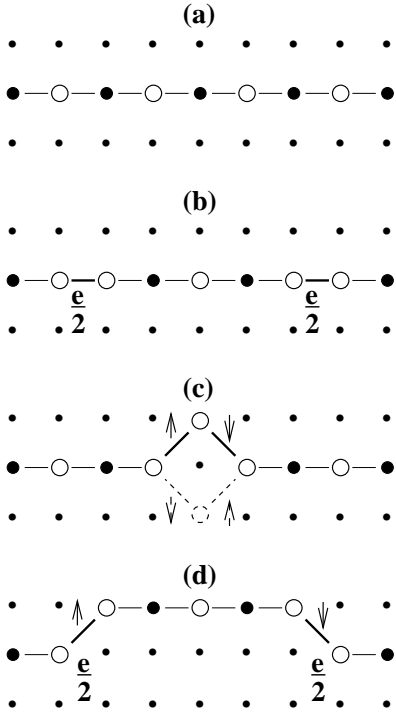}
\caption{Soliton dynamics in a strongly coupled doped $4k_F$ stripe. (a)
The reference state: localized stripe with $4k_F$ charge density 
wave. (b) If the stripe is rigid, the doped hole separates
in a left- and right moving soliton, both carrying half the electron
charge. (c) When the curvature energy becomes less than
the charge compressibility energy, the hole can escape `sideways'. (d)
As a result, the solitons now carry a transversal (step up/down)
flavor, which is like
a spin degree of freedom. Holes tunneling through the stripe 
fluctuate the transversal flavor, see (c).}  
\label{f1}
\end{figure}

The most elementary physical interpretation of the quantum 
sine Gordon model, Eq. (\ref{cdwlut}), actually corresponds with a free
string moving on a lattice: the field $\phi$ is the transversal
displacement ($z(l)$) at point $l$ of the string, while the cosine term
describes the lattice washboard on which the string moves ($\alpha \phi_c \rightarrow 2\pi z(l) / a$). The weak- and strong coupling limits are 
easily understood as a freely meandering string and one which is fully 
localized due to the lattice potential. As we discussed elsewhere in
detail\cite{eskes}, this notion is of relevance  in the context of 
transversally fluctuating  insulating stripes. In analogy with the
charge density wave problem, the relevant lattice scale dynamics is that of 
transversal solitons or `kinks'. Consider the vicinity of the string
delocalization
transition. Because of the domination of the lattice potential the string
tends to be localized on a particular lattice row $n$, and the exceptions
are where the string jumps to neighboring rows $n \pm 1$ (Fig. 1c,1d).
The origin of the  collective motions of the string lies in
the microscopic dynamics of these kink-excitations, while the tightly localized 
kinks of Fig. 1 represent a legitimate starting point, connected by adiabatic 
continuation to a class of string microscopies.

The existence of a localized stripe with internal $4k_F$ density
wave allows for a simple unification of the microscopic string-  
and internal charge dynamics.  Obviously, the fixed 1D electron 
trajectory assumed in the Luttinger liquid is no longer a given for 
electronic stripes. For a fixed trajectory,
it costs an energy equal to the jump in the thermodynamic potential $\delta
\mu$ to dope the charge density wave
with an additional carrier.
On the stripe, this (`longitudinal') energy cost can be reduced by letting
the charge escape `sideways', causing a transversal displacement,
at the expense of paying a curvature energy. Hence, when this curvature
energy becomes less than the energy cost associated with compressing the
charge, the doped holes will `carry a string fluctuation'.
In terms of the strong coupling kinks/solitons, the microscopic mechanism of 
transversal relaxation is obvious: {\em the doped hole corresponds
with a double kink in the string which is at the same time a 
soliton-antisoliton pair in the on-string charge density wave} -- see
Fig. 1. Starting from the CDW/localized string reference state, the
kinks and the solitons are the same objects, and this corresponds with
the strongest possible microscopic coupling between the on-string
metallicity and the string fluctuation.  
Due to the string fluctuation, the CDW solitons acquire
a {\em transversal flavour}: the (anti) soliton can move the string either
in an `upward' ($\uparrow$) or `downward' ($\downarrow$) direction 
(Fig. 1c/1d). This 
transversal freedom is like a $s=1/2$ iso-spin degree of freedom.
Since the CDW solitons can be described in terms of spinless fermions,
the string soliton dynamics relates to a {\em spinful} fermion system.
Since the string dynamics is like the spin dynamics in a standard 1D metal,
it follows that the separation of charge- and string dynamics is generic.

The qualitative nature
of the long wavelength physics can be deduced from the strong coupling
cartoon of Fig. 1, leaving the non-universal parameters of the theory
to be determined from a more realistic microscopic theory.  We
seek a generalization of the spinless fermion model, 
incorporating the string flavor in terms of isospin labels 
$\uparrow$ and $\downarrow$  attached to the fermions (see Fig. 1c,1d). 
As a first guess, one could take the spin-full
version of Eq. (\ref{cdwsol}) with a hard-core ($U \rightarrow \infty$)
condition:  the string flavor is conserved under the hopping of the
solitons. However, this
neglects the specifics of the transversal sector: (i) curvature energy
is associated with the order of the iso-spins. Obviously a $\uparrow
\downarrow$ isospin configuration of neighboring solitons involves
a different curvature energy than parallel configurations. These 
curvature energies can be absorbed in isospin-only Ising terms $\sim
S^z_i S^z_j$ $(\vec{S}_i = \sum_{\alpha \beta} c^{\dagger}_{i\alpha}
( \vec{\sigma} )_{\alpha \beta} c_{i\beta}$). (ii) The overall
transversal string displacement $u$ after arclength $r$
becomes ($a_0$ is the lattice constant),
\begin{equation}
\label{strfl}
 u(r) - u (0) = a_0 \int_0^r dx \sigma^z (x) 
\end{equation} 
where $\vec{\sigma} (x) = \sum_m \vec{S_n} \delta (x - x_m)$ ($x_m$ 
is the position of the m-th kink).
As long as this quantity is conserved the string remains localized.
Since $U \rightarrow \infty$ there is no kinetic exchange, and Ising
isospin terms do not fluctuate $u(l)$ either. In order to fluctuate
the string displacement the isospins should be exchanged and this is
possible if and only if two kinks recombine into a hole, because the hole can
tunnel through the string, see Fig. 1c. The simplicity of the argument is
deceptive: {\em this is an explicit realization 
of the idea of topological confinement}\cite{and1}.
Because of their topological nature, the kinks are strictly limited to 
the 1D string trajectory. In order to sweep the string through 2D
space, the kinks have to pair up in holes, because the latter
can propagate in 2D. 

In isospin language, the hole tunneling corresponds with spin-flip (XY) terms
$\sim S^+_i S^-_j + h. c. $. Notice that the energy barrier involves 
the {\em difference} in curvature energy and the charge-compression energy. 
This might well be a small number, and the hole-tunneling
rate can in principle be large. 
Assuming everything to be short ranged, we arrive at
the following model, in standard notation ($n = n_{\uparrow}+n_{\downarrow}$),
\begin{eqnarray}
\label{strferc}
H^{str}_0 & = & -t \sum_{n\sigma} ( c^{\dagger}_{n+1\sigma} c_{n \sigma} +
c^{\dagger}_{n \sigma} c_{n+1, \sigma} )  \nonumber \\
&  & + V \sum_n n_n n_{n+1} + U \sum_n n_{n\uparrow} n_{n\downarrow}  
 + J_{//} \sum_{<n m>} S^z_n S^z_m \nonumber \\
& & + { {J_{\perp}} \over 2} \sum_{<n m>} \; ( \; S^+_n S^-_m + S^-_n S^+_m \; )
\end{eqnarray}
$U$ should be taken to
infinity, while $t, V, J_{//}$ and $J_{\perp}$ parametrize the kink
kinetic energy, the `string neutral' kink repulsion, the curvature energy
and the hole tunneling rate, respectively. 

Although we are not aware of explicit calculations on this particular
model, the structure of the long wavelength dynamics can be deduced
directly\cite{repphys}.  When $J_{//} = J_{\perp}$,
Eq. (\ref{strferc}) is like the extended Hubbard model model with finite $U$,
at a low carrier density. The general case
$J_{//} \neq J_{\perp}$ corresponds with an interacting electron system 
with a spin-orbit coupling causing uniaxial spin anisotropy. The structure
of the long wavelength theory can be obtained from the weak-coupling limit,
and we refer to the extensive g-ology analysis including spin-orbit coupling
by Giamarchi and Schulz\cite{giasch}. Quite generally:  (i) we already
mentioned that charge and string flavor will separate always.
(ii) The charge dynamics is described by QSG,
 Eq. (\ref{cdwlut}), and away from the quarter-filled
point Umklapp becomes irrelevant. The charge dynamics at long wavelength
is therefore  described by free fields characterized by the fully
renormalized charge-velocity and -stiffness, $\tilde{v_c}$ and 
$\tilde{K_c}$ which will behave similarly as to the ones of the
extended Hubbard model in the strongly coupled regime. (iii) A crucial
observation is that the gross behavior in the string sector is determined
by the `isospin-only' problem\cite{giasch}. 

The isospin dependencies of the interactions are 
explicit in Eq. (\ref{strferc}), and the isospin-only problem is nothing
else than a XXZ problem with $S=1/2$, which has been solved a long time
ago\cite{lupesch}. 
If $-1 < J_{//}/J_{\perp}  < 1$, the Ising interaction is irrelevant
and the system falls in $XY$ universality, as described by free-field theory -
 the free string is
recovered. When  $|J_{//}| > |J_{\perp}|$ the Ising anisotropy takes over
and the string modes acquire a mass --  metallicity is a necessary
but insufficient condition for the string delocalization.
Physically, strings in this
regime are of the `disordered flat' variety\cite{eskes}. 
Although kinks proliferate and
delocalize, their internal string flavor (isospin) is ordered, as a compromise
between kinetic energy and lattice commensuration energy. 
The `ferromagnetic' case ($J_{//} < |J_{\perp}|$) corresponds with 
a `slanted' phase\cite{eskes}: the string is still localized, but it takes
some direction in space determined by the density of kinks. For  
$J_{//} > |J_{\perp}|$ the string is at average bond centered, and this
phase is related to the hidden order present in Haldane 
spin chains\cite{eskes}.
 
The most interesting phase is the delocalized string, and we will now show 
that the asymptotic structure of Luttinger liquid theory implies a rather
weak influence of the string-metallicity on the string fluctuation. 
A quantity of physical interest is the mean square transversal displacement
of the string\cite{zhvs}, using Eq. (\ref{strfl}),
\begin{equation}
\label{strdis}
\langle ( u(r)-u(0) )^2 \rangle = a_0^2\int_0^{r} dx dx' \langle \sigma^z (x) 
\sigma^z (x') \rangle
\end{equation}
The spin-spin correlation function of a one dimensional metal has the 
asymptote ($K_{\sigma}$ is the spin-stiffness),
\begin{equation}
\label{spincor}
\langle \sigma^z (x) \sigma^z (0) \rangle = { {C_1} \over {x^2}  } +
 { {C_2 cos(2k_F x)}  \over  { | x |^{\eta }} }
\end{equation}
where $\eta = K_{\sigma} + K_{\rho}$. Although
$\eta \geq 1$, it can be less than two and the
staggered component of the spin-spin correlator could dominate the behavior
of the string correlator, Eq. (\ref{strdis}). However, it is easy to see 
that in the additional integrations in Eq. (\ref{strdis}) the staggered
component behaves as if it falls off by one power more than $\eta$
($\int dx cos(2k_f x) / x^{\eta} \rightarrow \int dx 1/ x^{\eta +1}$). Since 
$\eta \geq 1$ it follows that the large $r$ asymptote of Eq. (\ref{strdis})
is governed by the uniform  component $\sim C_1$ in Eq.(\ref{spincor}).
Using that $\int_0^r dx dx' f( x- x') = \int_{-r}^r (2r - | x | ) f (x)$ and
the fact that $\int_{-\infty}^{\infty} dx 
\langle \sigma^z(x) \sigma^z(0) \rangle = 0$ it follows that the metallic
string behaves asymptotically as a free string\cite{zhvs},
\begin{equation}
\langle ( u(r)-u(0) )^2 \rangle = - 2 a^2_0 C_1 \ln ( r / r_c) + const.
\label{strlog}
\end{equation}
with a constant coming from short wavelength physics and
introducing a microscopic cut-off $r_c$.

Although not often discussed, the amplitude $C_1$ of the uniform component
of the spin-spin correlation is also in the metal entirely determined by 
the spin sector, which implies in the present context that the strength
of the string fluctuation is determined primarily by the transversal sector.
This can be easily understood from the insight by Schulz\cite{schulz}
 that the charge sector of
the Luttinger liquids is nothing else than a 1D harmonic (`floating') Wigner
crystal of (in our case) solitons. To every soliton a spin is attached 
and Schulz shows that by factorizing 
$\langle \sigma^z (z) \sigma^z (0) \rangle$ in a spin- and a charge 
correlator and treating the charge sector on the Gaussian level,
it follows that the exponent $\eta$ in the staggered component of Eq.
(\ref{spincor}) is the sum of the charge- and spin stiffnesses because
the spin system `rides' on the harmonically fluctuating charge solid.
Following the same alley, it is straightforward to show \cite{unpub}
that this charge fluctuation is invisible in the uniform correlations
responsible for the string delocalization.

We are now in the position to completely quantify Eq. (\ref{strlog}). Using
Haldane's expressions for the Luttinger liquid correlation 
functions\cite{hald}  and
realizing that the cut-off $r_c$ corresponds with the lattice constant $a$ of
the soliton Wigner crystal,
\begin{equation}
\langle ( u (r) - u ( 0 ) ) ^2 \rangle = { {a_0^2 K_{\sigma} } \over {2\pi^2} }
\ln ( r/a ) + const.
\label{strlog1}
\end{equation}
Let us now assume that the above model applies litterally 
to cuprates. Assuming that finite range string-string interactions are
unimportant, a measure for the importance of the single string quantum 
fluctuations is the quantum collision length $\xi_c$, obtained by demanding
that the r.m.s. displacement of a string becomes of order of the
mean string-string separation d ($\simeq 4a_0$)\cite{zhvs}: 
$\langle ( u (\xi_c) - u ( 0 ) ) ^2 \rangle = d^2$. Using that the
soliton lattice constant $a = a_0 / (8x - 1)$ in terms of the doping
density $x$, and the expression for the spin stiffness\cite{lupesch}
$K_{\sigma}^{-1} =  1/2 + (1/\pi) \arcsin ( J_{//}/J_{\perp} )$, we obtain
$\xi_c = a_0 / ( 8x-1) \exp[ (d/a_0)^2 [ \pi^2 + 2\pi \arcsin ( J_{//} /
J_{\perp} ) ].$
 The doping density only enters in the prefactor via the
trivial soliton-lattice constant rescaling, while $\xi_c$ depends 
exponentially on the stripe separation and the transversal scales.
Hence, the metallicity induced long-wavelength string 
fluctuations can only play a decisive role in the quantum melting of
the stripe phase if the factor in the exponent becomes of order unity.
Because of the various numerical factors, this only happens
if the string sector is very close to the `ferromagnetic' point
$J_{//}/ J_{\perp} \rightarrow -1$. It appears as very unlikely that
such a fine tuning occurs in cuprates and we conclude that on-string
metallicity is {\em not} an important factor for the quantum melting
of the stripe phase. 

In summary, we have adressed the problem of a lattice string which is 
internally a metal. Starting from specific microscopic assumption
inspired by cuprate stripes, we have shown that its long wavelength
dynamics is a straightforward generalization of the Luttinger liquid
where the usual theory has to be extended with a sector of 
transversal sound modes. Although intended as a demonstration of
the existence of a fixed-point (with likely a finite basin of attraction),
a litteral interpretation of the microscopic model shows that the
string fluctuation is quite insensitive to the internal metallicity.
As applied to cuprates, this observation offers a rational for the
surprising {\em insensitivity}  of the static stripe phases in e.g. LTT materials against stripe doping.      
 
Important discussions are acknowledged with H. J. Eskes in an early stage
of this research.

\end{multicols}

\end{document}